\documentclass[conference]{IEEEtran}
\IEEEoverridecommandlockouts
\usepackage{cite}
\usepackage{amsmath,amssymb,amsfonts}
\usepackage{algorithmic}
\usepackage{graphicx}
\usepackage{textcomp}
\usepackage{xcolor}
\usepackage[most]{tcolorbox}
\usepackage{enumitem} 
\usepackage{hyperref}

\usepackage{subfig}
\usepackage[export]{adjustbox} 

\def\BibTeX{{\rm B\kern-.05em{\sc i\kern-.025em b}\kern-.08em
    T\kern-.1667em\lower.7ex\hbox{E}\kern-.125emX}}
\begin{document}

\newcommand\copyrighttext{%
  \footnotesize \textcopyright~2025 IEEE. Personal use of this material is permitted.
Permission from IEEE must be obtained for all other uses, in any current
or future media, including reprinting/republishing this material for
advertising or promotional purposes, creating new collective works, for
resale or redistribution to servers or lists, or reuse of any copyrighted
component of this work in other works.\\[2pt]
This is the author's accepted version. The version of record appears in
the \emph{2025 IEEE International Conference on Data Mining Workshops
(ICDMW)}, DOI:  \href{https://doi.org/10.1109/ICDMW69685.2025.00148}{https://doi.org/10.1109/ICDMW69685.2025.00148}}
\newcommand\copyrightnotice{%
\begin{tikzpicture}[remember picture,overlay]
\node[anchor=south,yshift=10pt] at (current page.south) {\fbox{\parbox{\dimexpr\textwidth-\fboxsep-\fboxrule\relax}{\copyrighttext}}};
\end{tikzpicture}%
}

\title{Benchmarking LLMs for Threat Level Determination}

\author{\IEEEauthorblockN{1\textsuperscript{st} Han Wang}
\IEEEauthorblockA{\textit{Cybersecurity Unit} \\
\textit{RISE Research Institutes of Sweden}\\
Kista, Sweden \\
han.wang@ri.se}
\and
\IEEEauthorblockN{2\textsuperscript{nd} Murathan Kurfalı}
\IEEEauthorblockA{\textit{Intelligent systems Unit} \\
\textit{RISE Research Institutes of Sweden}\\
Kista, Sweden \\
murathan.kurfali@ri.se}
\and
\IEEEauthorblockN{3\textsuperscript{rd} Alfonso Iacovazzi}
\IEEEauthorblockA{\textit{Cybersecurity Unit} \\
\textit{RISE Research Institutes of Sweden}\\
Kista, Sweden \\
alfonso.iacovazzi@ri.se}
}

\maketitle
\copyrightnotice

\begin{abstract}
The fast progress of large language models (LLMs) opens new opportunities in the management of cyber threat intelligence, but their reliability for operational tasks remains unclear. In this work, we benchmark LLMs on the task of threat level determination. First, we construct a curated dataset derived from publicly available MISP OSINT feeds. Next, we design a tailored prompt to systematically compare eight different LLMs under zero-shot conditions. Finally, we apply supervised fine-tuning on each model and perform a comparative analysis between baseline and fine-tuned versions. Our results show that zero-shot models achieve weak performance, with limited ability to correctly assign threat levels. Fine-tuned models, however, demonstrate substantial improvements, reaching F1 scores between 0.40 and 0.58 depending on the base architecture. Despite this progress, the performance is still low for practical deployment, highlighting the need for additional research on data quality, model adaptation, and domain-specific tuning. 
\end{abstract}


\section{Introduction}
As cyber attacks become more frequent and sophisticated, organizations increasingly struggle to maintain a robust cybersecurity posture. Cyber Threat Intelligence (CTI) is essential in helping organizations detect and prevent potential threats by offering critical insights into vulnerabilities and attack vectors. These insights support informed risk mitigation decisions. However, traditional CTI management methods typically rely heavily on security experts, making them time-consuming, prone to errors, and demanding in terms of resources \cite{Iacovazzi2023}. This highlights the urgent need for more streamlined and effective approaches. CTI is often gathered from diverse sources and presented in various formats, resulting in data that is inconsistent and difficult to standardize, integrate, and analyze. Moreover, CTI data is frequently incomplete, lacking key contextual details such as the origin, intent, or severity of threats, which undermines confidence in its accuracy and can lead to flawed decision-making.

Threat management and sharing platforms help process, analyze, and share threat details. For example, Malware Information Sharing Platform (MISP) is an open-sourced and widely used by research organizations and private companies to exchange threat intelligence \cite{misp-paper}. Such platforms support the sharing, storing, and analyzing of diverse threat data, ranging from fraudulent banking transactions to unusual network activities. They also enable cybersecurity professionals to assign a threat level to each event or incident, indicating its severity and urgency. A high threat level typically corresponds to sophisticated attacks such as advanced persistent threats (APTs) or zero-day exploits, which could be very harmful and require immediate attention. In contrast, a lower threat level is usually associated with indicators of compromise (IoCs) linked to common malware with limited impact and reduced urgency. 

Assigning a threat level to threat events or incidents enables cybersecurity professionals to respond with greater precision and timeliness, thereby reducing the likelihood of misjudgments. Nevertheless, threat levels are usually determined manually, a process that is both time-intensive and susceptible to human error. Therefore, the automation of threat level determination is critical for enhancing the efficiency and reliability of CTI processing and analysis.

The advent of Large Language Models (LLMs) offers significant potential to improve the processing and analysis of CTI derived from highly heterogeneous data sources. A growing number of studies have begun to explore the applicability of LLMs to CTI-related tasks. However, to the best of our knowledge, no prior work has examined how LLMs perform in determining the threat level of individual threat incidents or events. While related studies have investigated the classification of Common Vulnerabilities and Exposures (CVEs) using the Common Vulnerability Scoring System (CVSS) metrics, these metrics are designed explicitly for vulnerabilities \cite{ghosh2025,marchiori2025,wang2025cve}. In contrast, the threat level determined for CTI exchange data is usually broader in scope and more heterogeneous, as incidents often include confirmed IoCs that encompass not only vulnerabilities but also zero-day exploits, APTs, malware, and other forms of malicious activity. Wang et al. \cite{wang2024} represent the first study to address the ranking of IoCs from MISP through a learning-to-rank algorithm; nevertheless, their approach does not incorporate LLMs.

To bridge this gap, we introduce a benchmark specifically designed to evaluate the effectiveness of LLMs in determining threat levels within the context of CTI automation for management and analysis. The benchmark dataset is derived from the MISP Open Source Intelligence (OSINT) feed. It is crafted through processes of simplification and normalization to ensure consistency, reduce noise, and facilitate model interpretability. Our evaluation framework follows a two-stage approach. First, we employ zero-shot prompting to examine the extent of LLMs’ inherent knowledge regarding assessing threat levels in threat data. Subsequently, we fine-tune the models on the dataset to systematically compare their performance, thereby providing insights into both their out-of-the-box capabilities and adaptability through task-specific training. We summarize our contributions in this paper as follows:
\begin{itemize}
\item To the best of our knowledge, this is the first work that assesses LLMs' performance on threat level determination. In doing so, it addresses a gap in the annotation of threat events and contributes to reducing reliance on manual human intervention.
\item We construct a structured and pre-processed dataset\footnote{The data preparation scripts, as well as the snapshot used in the current experiments, can be found here: \url{https://github.com/NLP-RISE/cti-threat-level-benchmark}} from MISP OSINT feeds, applying schema reduction, normalization and filtering. We also develop a tailored prompt for the task to benchmark it on LLMs.
\item We conduct supervised fine-tuning of the LLMs and systematically compare the impact of the input representation (original JSON vs structured more concise textual representations), showing that both fine-tuning and the choice of the input representation have a substantial effect on the outcome.
\item We aim our benchmark and experiments to serve as a foundation for future work on LLM-based threat level classification models. 
\end{itemize}

The rest of the paper is organized as follows. Section \ref{sec:relatedwork} first summarizes related work. Section \ref{sec:benchmark} explains the design for the threat level determination task and the crafted dataset. Section \ref{sec:approach} introduces the proposed evaluation framework for benchmarking LLMs. Section \ref{sec:evaluation} summarizes the experiments from different setups. In Section \ref{sec:discussion}, we discuss the experimental results and highlight the potential challenges. Finally, Section \ref{sec:conclusion} draws the conclusions and the future work.

\section{Related Works}
\label{sec:relatedwork}
LLMs are increasingly being applied to a variety of CTI tasks. Alves et al. \cite{alves2022leveraging} propose an approach for the automatic extraction of Tactics, Techniques, and Procedures (TTPs) from unstructured textual sources using BERT-based language models. Their study presents a comparative evaluation of several BERT variants on the MITRE ATT\&CK cyber threat report dataset \cite{strom2018mitre}, demonstrating that these models consistently outperform a baseline method that employs TF-IDF features with a Linear Regression classifier. Other works leverage LLMs to extract meaningful information and construct knowledge graphs for CTI \cite{WANG2024103824, HU2024103999, cheng2025ctinexusautomaticcyberthreat, 10628558}. For example, Hu et al. \cite{HU2024103999} applied few-shot learning and fine-tuning to both named entity recognition and MITRE’s Tactics, Techniques, and Procedures classification, achieving high accuracy, while Cheng et al. \cite{cheng2025ctinexusautomaticcyberthreat} proposed automating knowledge extraction using LLMs. Xu et al. \cite{xu2024} extended this line of research by extracting attack-level intelligence, incorporating implementation procedures and contextual reasoning to identify techniques and tactics, and generating procedures aligned with the MITRE ATT\&CK framework. Paul et al. \cite{paul2025} integrated Retrieval-Augmented Generation (RAG) with threat intelligence to enable real-time analysis, and Baral et al. \cite{11059476} advanced this direction further by developing a reasoning-and-acting agent powered by LLMs, capable of autonomously mitigating cyberattacks in real time. Collectively, these studies reflect an emerging trend in CTI research, where the role of LLMs is expanding from basic information extraction toward higher-level reasoning, contextual understanding, and even autonomous cyber defense.

Related research has explored using LLMs for determining CVSS metrics for CVEs \cite{ghosh2025,marchiori2025,wang2025cve}. For example, Marchiori et al. \cite{marchiori2025} investigate the capability of LLMs to generate CVSS scores for newly disclosed vulnerabilities and benchmark their performance against embedding-based approaches. Their work also examines prompt engineering strategies to improve accuracy and evaluates the potential of text embedding models and machine learning techniques for score generation. 

An emerging line of research focuses on creating robust benchmarks designed to systematically assess the capabilities and limitations of LLMs within the context of CTI \cite{alam2024ctibench, zhu2025cvebench, ji2024sevenllm, yong2025}. Unlike general-purpose language evaluation frameworks, these benchmarks emphasize the applied dimensions of CTI, including threat identification, attribution of threat actors, and analysis of attack methodologies. For instance, CTIBench \cite{alam2024ctibench} introduces a suite of tasks, CTI-MCQ, CTI-RCM, CTI-VSP, and CTI-TAA, intended to evaluate an LLM’s domain-specific comprehension and problem-solving proficiency in CTI. On the other hand, CVE-Bench is specifically developed to evaluate the ability of LLMs to identify and analyze critical CVEs \cite{zhu2025cvebench}. Such benchmarking works are essential for ensuring that LLM evaluation accurately reflects the complexities and realities of real-world cybersecurity threats. 

While these CTI benchmarks, including both CTIBench and CVE-Bench, provide valuable frameworks for assessing LLMs' performance on core intelligence tasks, they do not explicitly address the operational challenge of threat-level determination. Unlike CVSS scoring, which is a standardized severity assessment tied to individual software vulnerabilities, threat-level determination requires a broader contextual judgment that integrates heterogeneous indicators (e.g., attack patterns, actor intent, and observed impact) to classify the overall seriousness of a threat. Similarly, the tasks in CTIBench, such as multiple-choice comprehension, actor attribution, or vulnerability summarization, evaluate specific reasoning abilities, but they are not designed to evaluate whether LLMs can synthesize diverse alerts into actionable severity assessments. Our work, therefore, complements these benchmarks by focusing on the unique problem of threat-level determination, a task that lies at the intersection of situational awareness and decision support, and remains underexplored in current LLM evaluations.


\section{Benchmark design} 
\label{sec:benchmark}
In this section, we describe the designed task for threat level determination and the construction of the dataset used in our study. Our aim is to build a structured and reproducible corpus of threat intelligence events derived from MISP data feeds.

A CTI workflow is, in practice, the end-to-end process of collecting, processing, analyzing, and disseminating cyber threat information to transform raw data into actionable intelligence. The benchmark for threat-level determination designed in this work aims to automate the processing and analysis phases, thereby reducing human intervention while accelerating and strengthening overall CTI management. Once a CTI sharing platform such as MISP receives a series of incidents, the designed task can automatically assign a reliable threat level, which may then serve as a reference point for further validation by cybersecurity experts if necessary. 

To better reflect real-world cases, we collected raw CTI data directly from publicly available MISP OSINT feeds, provided by CIRCL\footnote{\url{https://www.circl.lu/doc/misp/feed-osint/}}  and Botvrij.eu.\footnote{\url{https://www.botvrij.eu/data/feed-osint/}}
We use only MISP OSINT feeds because, unlike most public CTI datasets, they provide manually assigned threat levels in addition to indicators. Crucially, these labels of the threat level are assigned by cybersecurity experts from diverse organizations, reflecting real-world usage of these OSINT feeds and offering an accurate representation of current cybersecurity threats. This is critical for our work, as it enables us to assess how effectively models can determine the appropriate threat level in real-world contexts.

The Computer Incident Response Center Luxembourg (CIRCL) serves as one of the principal contributors to the MISP project, actively supporting its development and operational use within the cybersecurity community. As part of its ongoing efforts, CIRCL maintains and disseminates an up-to-date OSINT feed, which provides high-value CTI to practitioners and researchers. Building on this resource, the CIRCL dataset was compiled by systematically collecting IoCs from the OSINT feed, offering a representative corpus for analyzing real-world threat intelligence. On the other hand, Botvrij.eu offers a variety of open-source IoCs that can be integrated into security devices to detect potential malicious activities. The data is collected from open-source information feeds --such as blog posts and PDF reports --and subsequently consolidated into structured datasets. 

Our snapshot, created on 18 August 2025, included 2005 MISP events. To transform this raw collection into a dataset suitable for experiments with LLMs, we applied a multi-stage processing pipeline, described in the rest of the section.

\subsection{Dataset extraction and curation}
The raw MISP feeds are quite heterogeneous. The MISP format structure is a well-defined JSON format and strongly related to how data is processed. The terminology in MISP can be categorized into two classes: the data layer and the context layer. The former includes all the terms related to how the information is defined in MISP, while the latter refers to the relationship between different clusters of information. The data layer contains \textit{Event}: The aggregation of contextually linked information, represented as attributes and objects; \textit{Attribute}: An individual data point, which can be an indicator or supporting data; \textit{Object}: The custom template for attributes. On the other hand, the context layer includes \textit{Tags}: The labels attached to events/attributes from MISP taxonomies; \textit{Galaxy-clusters}: The knowledge base item used to label events/attributes and comes from Galaxy; and \textit{Cluster relationship}: The relationship between Galaxy clusters.

In our study, each data sample is treated as a single threat event. The size of these events varies substantially, ranging from only a few attributes to several hundred. Many instances include redundant or non-informative elements for threat-level prediction, such as repeated entries or technical fields with limited semantic value (e.g., lengthy cryptographic hash values). To mitigate these issues, we implemented a strict schema-reduction and normalization procedure.

At the \textit{event} level, we retained a compact set of descriptors, including the date, a short textual description (\texttt{info}), timestamps, and the publication flag. We also preserved \textit{tags} and \textit{Galaxy clusters}, but restricted them to their names only, omitting stylistic details and sharing-related metadata. For \textit{attributes}, which typically contain concrete indicators of compromise (e.g., domain names, IP addresses, or CVE identifiers), we retained the type, value, category, timestamp, analyst comments, and the \texttt{to\_ids} flag indicating whether the attribute is considered actionable. Cryptographic hash values (e.g., SHA-256 or MD5) were replaced with placeholders (e.g., \texttt{<sha256>}), as their actual values do not contribute to threat-level prediction and would unnecessarily increase token counts. With respect to \textit{objects}, we preserved fields such as description, comments, meta-category, name, timestamp, and their associated \textit{attributes}.

This procedure eliminates low-level technical details while retaining the contextual information and indicators most relevant for assessing the severity of an event. At this stage, we produced an adapted version of the JSON representation in which no events were discarded; rather, their structure was streamlined and simplified.

\subsection{Filtering and Splitting}
After normalization, we introduced an additional filtering step to ensure the dataset’s suitability for our task. First, we removed events with missing or invalid threat levels, such as those labeled as ``Undefined.'' Second, we applied a length filter, since we observed that certain events were unusually long (e.g. exceeding millions of tokens) and would, otherwise, be truncated at inference time due to computational costs. To address this, we used the Llama 3.1 tokenizer as a representative tokenizer. Each event (in its adapted JSON form) was tokenized, and examples that exceeded a threshold once the prompt and a small generation buffer were added were discarded.

Concretely, we assumed a fixed maximum context of 8192 tokens. We reserved 400 tokens for the prompt and 100 tokens for the model’s output. To account for tokenizer variance, we added a 10\% buffer (because different LLMs tend to yield different tokenizations). This resulted in a practical threshold of approximately 6,500 tokens. 

Starting from 2005 raw events in our snapshot, we excluded 73 ``Undefined'' cases and an additional 310 events that exceeded the length threshold. The final labeled dataset, therefore, contains 1,622 events. We then performed a stratified split to preserve class proportions. The distributions are shown in Table~\ref{tab:data_stats}.  

\begin{table}[t]
\centering
\small
\caption{Threat-level distributions before and after filtering.}
\begin{tabular}{lc|rr}
\hline
Threat level & Raw & \multicolumn{2}{c}{Filtered} \\
\cline{3-4}
 &  & Test & Train \\
\hline
High & 196 (9.8\%) & 47 (9.7\%) & 111 (9.8\%) \\
Medium & 432 (21.5\%) & 107 (22.0\%) & 251 (22.1\%) \\
Low & 1304 (65.0\%) & 332 (68.3\%) & 774 (68.1\%) \\
Undefined & 73 (3.6\%) & -- & -- \\
\hline
Total & 2005 (100.0\%) & 486 (100.0\%) & 1136 (100.0\%) \\
\hline
\end{tabular}
\label{tab:data_stats}
\end{table}

 \begin{figure}[t]
    \centering
    \includegraphics[width=0.5\textwidth]{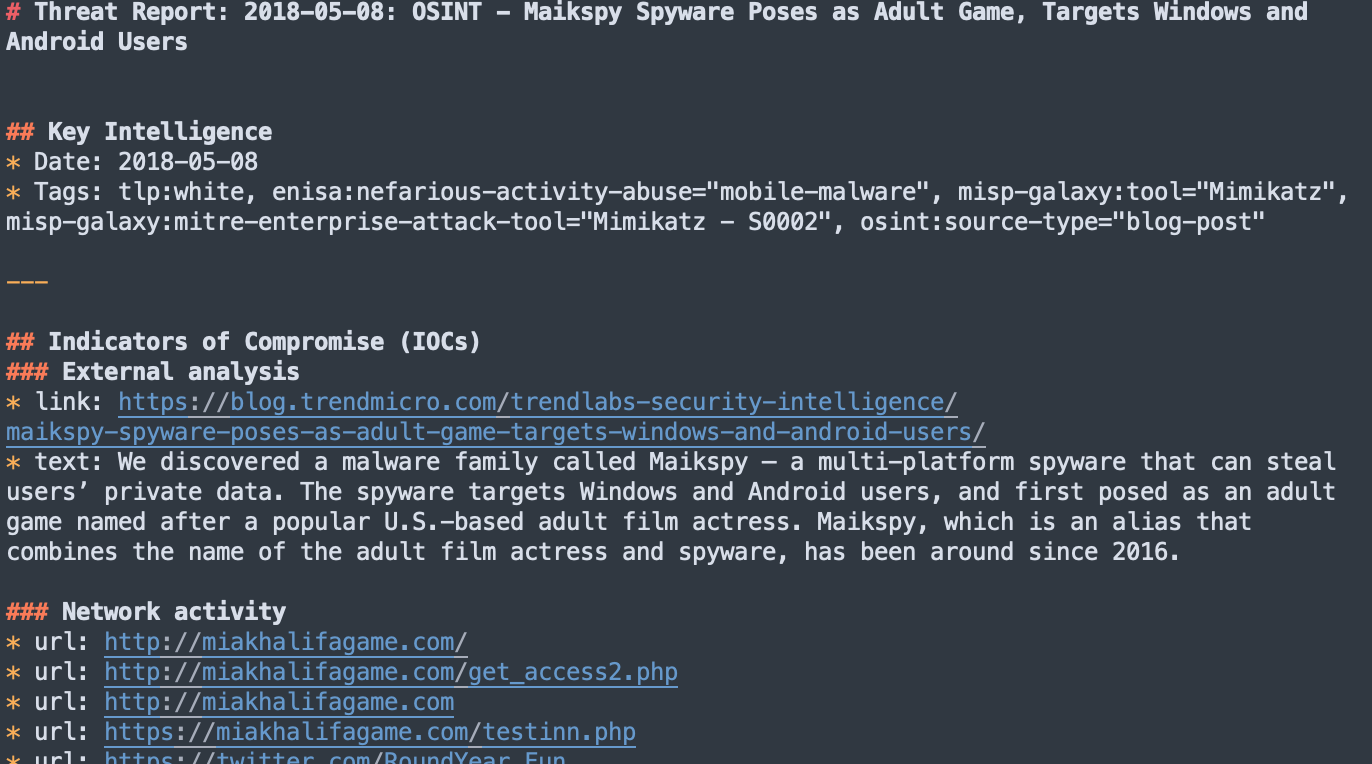} 
    \caption{Example for textual representation.}
    \label{fig:md}
\end{figure}

\subsection{Textual Representation}
In addition to the adapted JSON representation, we generated a parallel, more human-readable textual representation. This transformation converts the JSON files into structured text while preserving their contents. The new representation organizes core event details, tags, indicators of compromise, and objects into a layout with clear headings and bullet points. A sample of textual representation of an actual MISP event from our dataset is given in Figure \ref{fig:md}.

This representation does not introduce any new information beyond what is available in the adapted JSON, but it considerably decreases the token count by removing JSON artifacts such as braces and quotation marks. We hypothesize that this more compact and natural text format may improve the performance of LLMs. Therefore, we produced a textual version of each JSON in our training and test set to systematically evaluate whether the input representation has an impact on threat-level classification performance.

\begin{figure}[t]
    \centering
    \includegraphics[width=0.5\textwidth]{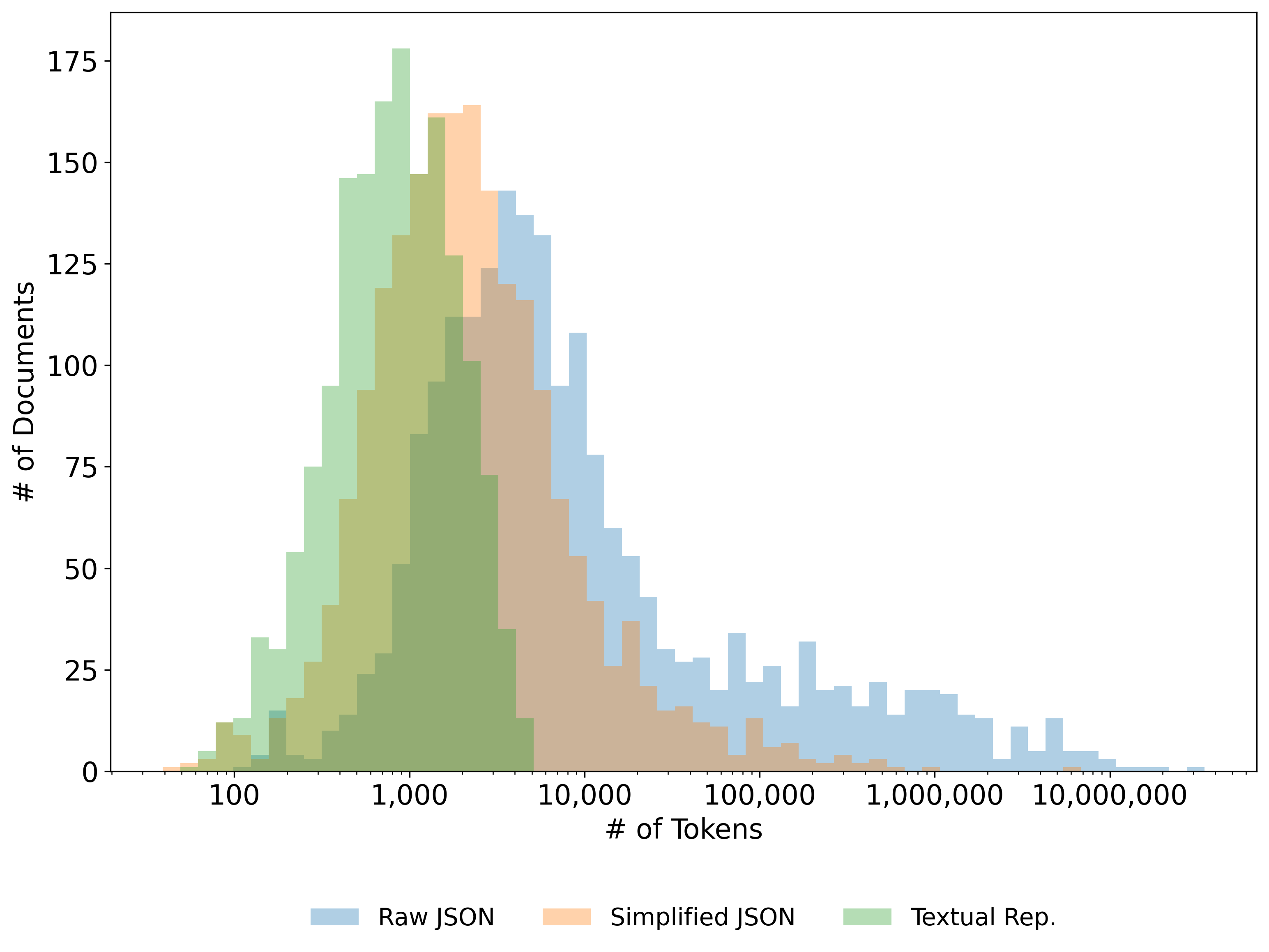} 
    \caption{Token length distributions for different input representations. The original MISP JSON files contain a long tail of very large events, with some over one million tokens. Our pre-processing pipeline substantially reduces this skew, resulting in more compact inputs.} 
    \label{fig:histrogram}
\end{figure}

To further illustrate the effects of our pre-processing pipeline, we provide the token length distributions for original MISP files, their simplified version, and the textual representation in Figure \ref{fig:histrogram}. The original MISP files, denoted as \textit{Orig. MISP}, exhibits a very long tail with some events exceeding millions of tokens. Simplification substantially reduces this tail, concentrating most events around 5000 tokens. The textual representation is the most compact, with the bulk of events between a few hundred and a few thousand tokens.

\section{Methodology}
\label{sec:approach}
\subsection{Zero-shot Propmting}
We first evaluate LLMs in a zero-shot setting. To this end, we construct a dedicated prompt that explicitly defines the three severity levels (high, medium, and low). Initial experiments showed that smaller models, especially, often produced verbose or inconsistent responses, making it difficult to extract the predicted label automatically. To overcome this problem, we require all models to follow a strict but simple JSON output format, containing only a single key \texttt{"label"} with one of the three valid values. This constraint ensures reliable automatic evaluation across models. The exact prompt used in our experiments is shown below.

\begin{tcolorbox}[
valign=top,
    colback=black!7!white, 
    colframe=black!75!black, 
    fonttitle=\bfseries,  
    title=Prompt   
]

Use these MISP-style definitions:
\begin{description}[
    style=unboxed,
    leftmargin=1.6cm,   
    labelwidth=1.4cm  
]
    \item[high:] Confirmed or highly credible threat; could be sophisticated; advanced persistent threats and zero-day exploits; severe impact likely; urgent to be mitigated.
    \item[medium:] Credible indicators of compromise or partial evidence; could be potential advanced persistent threats; meaningful impact possible.
    \item[low:] Weak indicators of compromise or single threat incident; could be general malware or reconnaissance activity; limited impact expected.
\end{description}
\noindent \textbf{OUTPUT FORMAT:}
\begin{itemize}
    \item Return ONLY a single valid JSON object (no Markdown, no extra text).
    \item Exactly one key: \texttt{"label"}.
    \item The value for \texttt{"label"} must be exactly one of: \texttt{high}, \texttt{medium}, or \texttt{low}.
\end{itemize}

\noindent \textbf{Example:}
\begin{verbatim}
{"label":"high"}
\end{verbatim}

\hrulefill

\noindent Now reply with the JSON object only for the following event. \\

\noindent \textbf{Event:} 
\textit{\{$<$contents of the MISP file$>$\}} \quad 
\end{tcolorbox}

We would like to note that we deliberately refrain from performing few-shot experiments. As discussed above, MISP events are often lengthy, and adding even a small number of demonstration examples would significantly reduce the effective context available to the model and risk truncating the input.

\subsection{Supervised Fine-Tuning}
We fine-tune models on the training split of our dataset using low-rank adapters (LoRA) \cite{hu2022lora}. Each training instance is formatted as a short conversation (system + user) with the same instruction as in the zero-shot setting; the target assistant turn is the label-only JSON. Training applies \emph{response-only loss}, meaning that optimization is performed exclusively on the assistant output. The overall training configuration is summarized in Table~\ref{tab:ft}. 

We explore different LoRA configurations by varying the rank ($r \in \{8,16\}$) and scaling factor ($\alpha \in \{32,64\}$), following common practice.\footnote{A preliminary experiment with larger values ($r=32$, $\alpha=128$) led to degraded performance.} Additionally, we compare training on the full dataset with a downsampled variant, where the number of instances per label is balanced by randomly reducing all classes to match the frequency of the least common label, \textit{High}.

\begin{table}[t]
\centering
\small
\caption{Fine-tuning hyper-parameters}
\label{tab:ft}
\begin{tabular}{l l}
\hline
Component & Setting \\
\hline
Max sequence length & 8192 \\
Optimizer & AdamW with cosine decay, 3\% warm-up \\
Learning rate & $2\times 10^{-4}$  \\
Batch size & 16 (via accumulation) \\
Epochs & 4 \\
LoRA Rank & \{8, 16\} \\
LoRA Alpha & \{32, 64\} \\
\hline
\end{tabular}
\end{table}

\subsection{Experimental Setup}

We evaluate a diverse set of instruction-tuned, open-source LLMs on our dataset. Table~\ref{tab:models} provides abbreviated model names alongside their full Hugging Face identifiers.\footnote{https://huggingface.co/} To ensure reproducibility and reduce output variability, we set the temperature at 0 across all experiments.

Performance is measured using both accuracy and macro-F1. While accuracy offers a straightforward indicator of overall correctness, macro-F1 is more meaningful given the class imbalance in our dataset. Macro-F1 gives equal weight to all classes and serves as our main metric for fair comparisons between models.

\begin{table*}[t]
\centering
\small
\caption{Models evaluated in our experiments. The model sizes are highlighted in bold.}
\label{tab:models}
\begin{tabular}{ll|ll}
\hline
Model & Hugging Face identifier & Model & Hugging Face identifier \\
\hline
Command-R+ (\textbf{111B}) & CohereLabs/c4ai-command-r-plus-4bit 
& Gemma-3-\textbf{1B}-IT & google/gemma-3-1b-it \\
GPT-OSS-\textbf{20B} & openai/gpt-oss-20b 

& Gemma-3-\textbf{12B}-IT & google/gemma-3-12b-it  \\
Llama-3.1-\textbf{8B} & meta-llama/Llama-3.1-8B-Instruct

& Gemma-3-\textbf{27B}-IT & google/gemma-3-27b-it \\

Llama-3.3-\textbf{70B} & meta-llama/Llama-3.3-70B-Instruct
& Ministral-\textbf{8B}-2410 & mistralai/Ministral-8B-Instruct-2410 \\
Qwen3-\textbf{4B}-2507 & Qwen/Qwen3-4B-Instruct-2507 & & \\
\hline
\end{tabular}
\end{table*}

All experiments were implemented through the transformers library \cite{wolf2020transformers} and executed on a single NVIDIA H100 (92 GB) GPU. A single fine-tuning run (4 epochs) completed within an hour, depending on model size (1B to 111B). The aggregate compute budget across all models is approximately 25 GPU-hours, including all test-set evaluations.

\section{Results}
\label{sec:evaluation}

Table~\ref{tab:leaderboard_base} summarizes zero-shot performance. Overall, results indicate that current instruction-tuned LLMs struggle to reliably infer the threat level of MISP events without task-specific adaptation. 
The majority baseline, which is a dummy classifier that always predicts the most frequent label, \textit{Low}, yields 0.683 accuracy and 0.271 macro-F1 on our test set. Only two models exceed this threshold: Command-R+ (111B), which achieves a macro-F1 of 0.313 with structured text (Text) input, and Llama-3.1-8B, reaching 0.277 under the same setting.

On the other hand, supervised fine-tuning has a notable effect on the performance. Table~\ref{tab:leaderboard_finetuned_64} presents results for models fine-tuned using LoRA with rank 32 and scaling factor $\alpha=64$. All models were trained on a balanced subset of the dataset (111 events per threat level) because full-dataset fine-tuning, as further discussed in Section~\ref{sec:how-to-finetune}, led to degraded performance, due to overfitting on the dominant \textit{Low} class.

Fine-tuning with the balanced training set leads to substantial improvements across all models, with macro-F1 scores increasing by 20–35 points relative to zero-shot baselines. The best-performing model is Llama-3.3-70B, reaching 0.574 macro-F1 and 0.704 accuracy with textual input, followed by Command-R+ (0.554 F1, 0.698 acc) and Llama-3.1-8B (0.520 F1, 0.675 acc). Even smaller models such as Gemma-3-12B and Gemma-3-1B achieve competitive performance (0.501 and 0.409 macro-F1, respectively).

\begin{table}[t]
\centering
\caption{Threat-level classification for base (non–fine-tuned) models. The best per model is chosen separately for JSON and text inputs.}
\label{tab:leaderboard_base}
\begin{tabular}{lcccc}
\hline \hline
 & \multicolumn{2}{c}{JSON} & \multicolumn{2}{c}{Text} \\
Model & Acc & F1 & Acc & F1 \\
\hline
Majority baseline &0.683 & 0.271&0.683 & 0.271\\ \hline
Command-R+ (111B) & 0.208 & 0.207 & 0.317 & 0.313 \\
Llama-3.1-8B & 0.249 & 0.228 & 0.315 & 0.277 \\
Llama-3.3-70B & 0.237 & 0.240 & 0.265 & 0.267 \\
Gemma-3-27B-IT & 0.030 & 0.019 & 0.230 & 0.243 \\
Gemma-3-12B-IT & 0.045 & 0.029 & 0.208 & 0.213 \\
Qwen3-4B-2507 & 0.177 & 0.178 & 0.185 & 0.191 \\
GPT-OSS-20B & 0.138 & 0.130 & 0.192 & 0.190 \\
Ministral-8B-2410 & 0.126 & 0.122 & 0.146 & 0.146 \\
Gemma-3-4B-IT & 0.025 & 0.016 & 0.146 & 0.146 \\
Gemma-3-1B-IT & 0.005 & 0.003 & 0.095 & 0.059 \\
\hline \hline
\end{tabular}
\end{table}

\begin{table}[t]
\centering
\caption{Threat-level classification for fine-tuned models (LoRA 32-64). Best per model is chosen separately for both input representations.}
\label{tab:leaderboard_finetuned_64}
\begin{tabular}{lcccc}
\hline \hline
 & \multicolumn{2}{c}{JSON} & \multicolumn{2}{c}{Text} \\
Model & Acc & F1 & Acc & F1 \\
\hline
Llama-3.3-70B & 0.669 & 0.546 & 0.704 & 0.574 \\
Command-R+ (111B) & 0.689 & 0.517 & 0.698 & 0.554 \\
Llama-3.1-8B & 0.640 & 0.480 & 0.675 & 0.520 \\
Gemma-3-12B-IT & 0.846 & 0.305 & 0.669 & 0.501 \\
Gemma-3-1B-IT & 0.741 & 0.284 & 0.504 & 0.409 \\
\hline \hline
\end{tabular}
\end{table}

\begin{figure*}[t]
\centering

\subfloat[Llama-3.1-8B (Zero-shot)]{
  \includegraphics[width=0.24\textwidth]{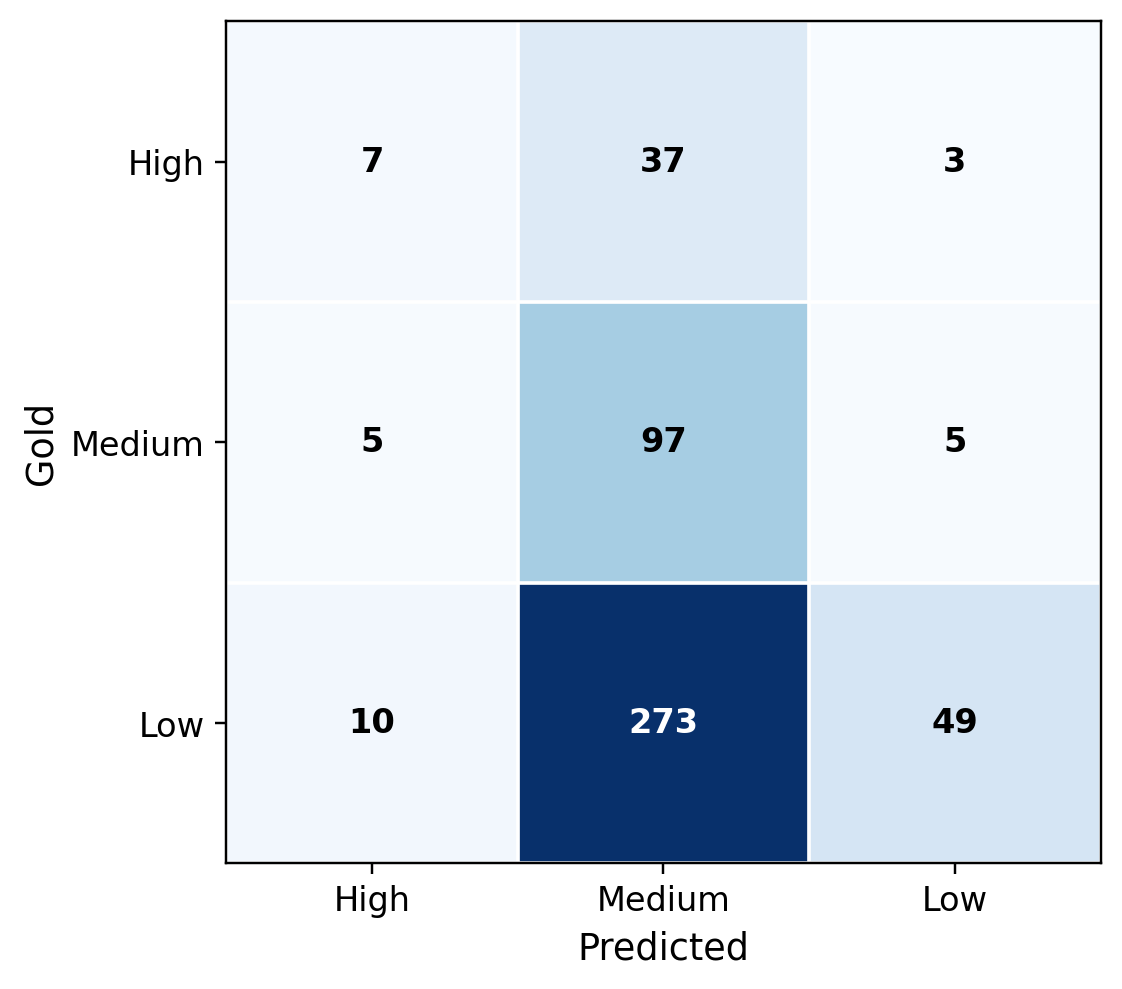}
}
\subfloat[Llama-3.1-8B (Fine-tuned)]{
  \includegraphics[width=0.24\textwidth]{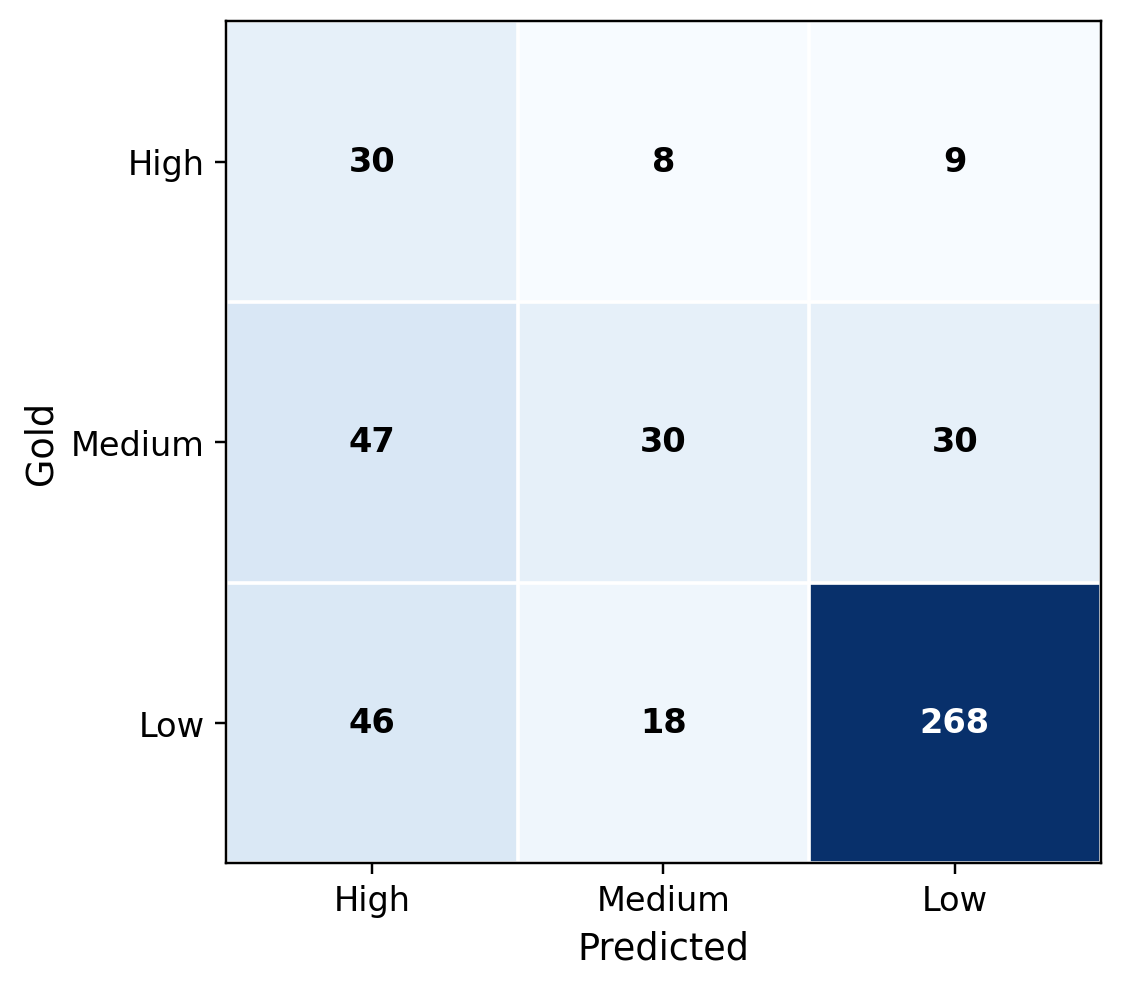}
}
\subfloat[Llama-3.3-70B (Zero-shot)]{
  \includegraphics[width=0.24\textwidth]{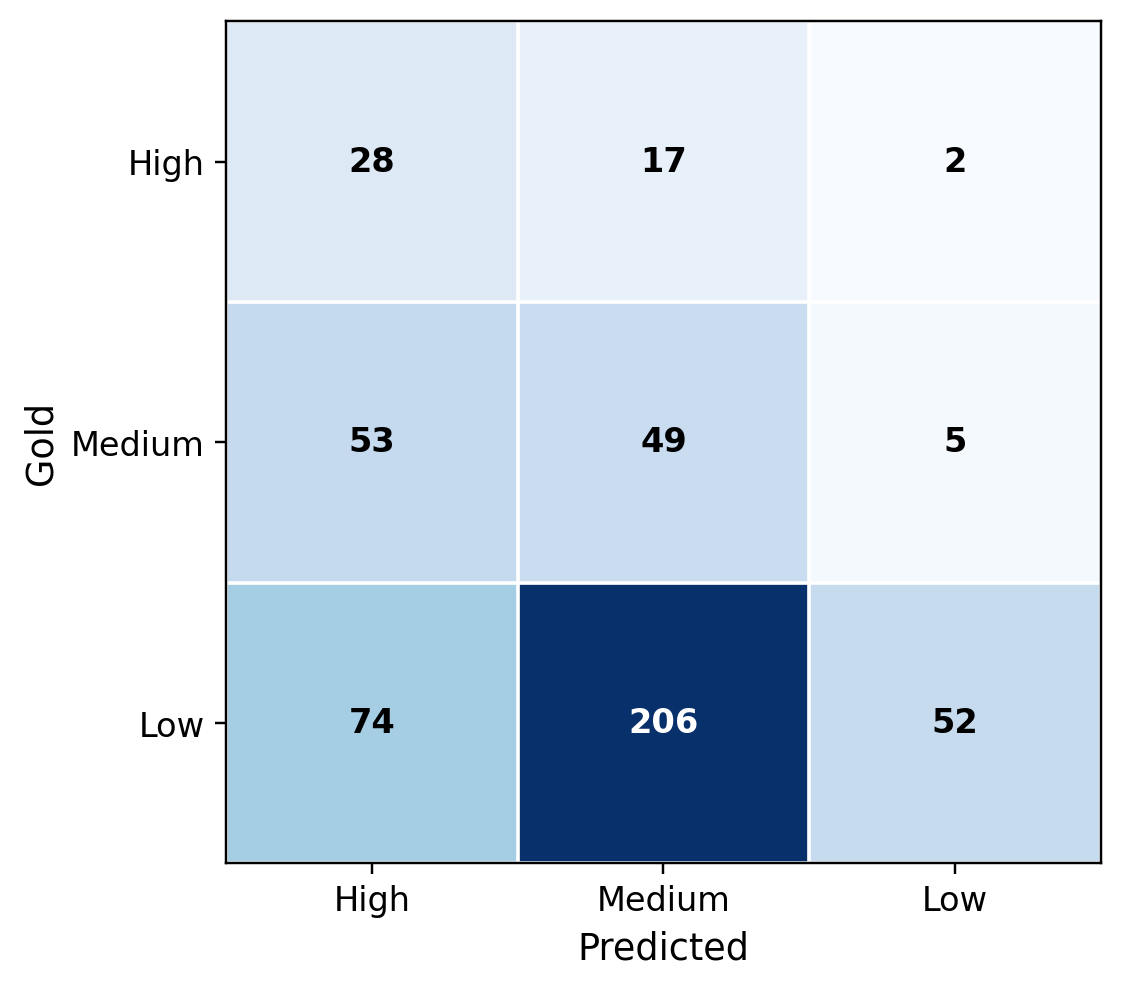}
}
\subfloat[Llama-3.3-70B (Fine-tuned)]{
  \includegraphics[width=0.24\textwidth]{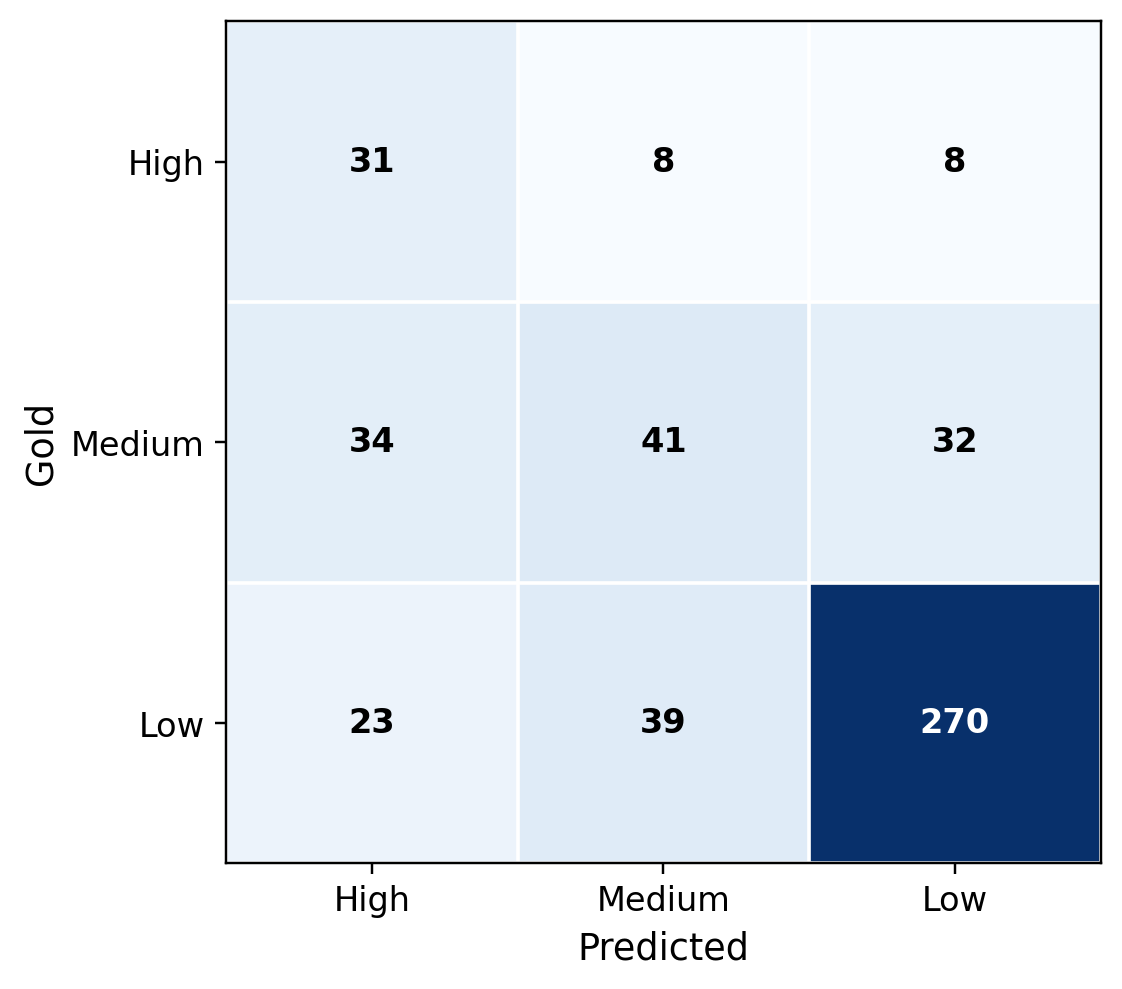}
}
\\
\subfloat[Gemma-1B (Zero-shot)]{
  \includegraphics[width=0.24\textwidth]{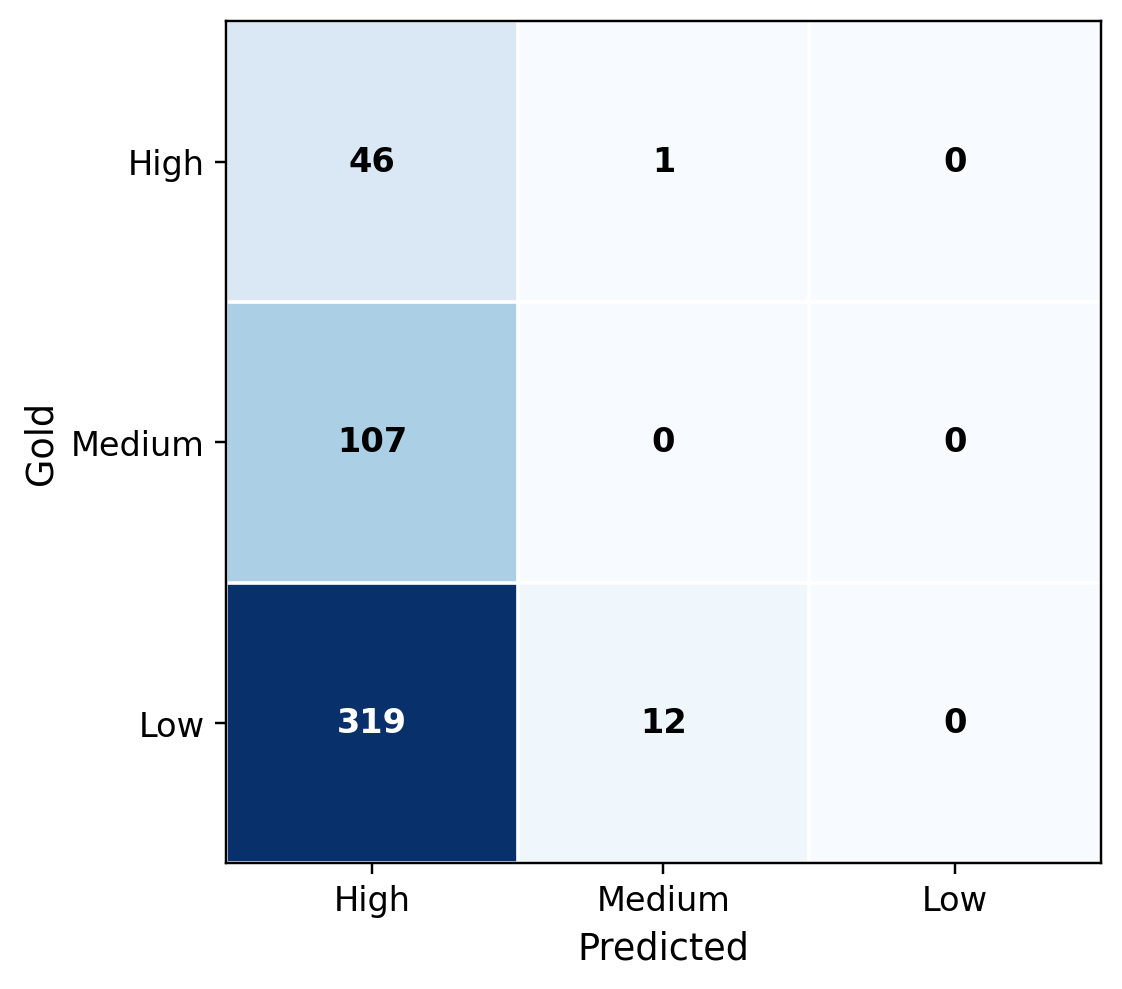}
}
\subfloat[Gemma-1B (Fine-tuned)]{
  \includegraphics[width=0.24\textwidth]{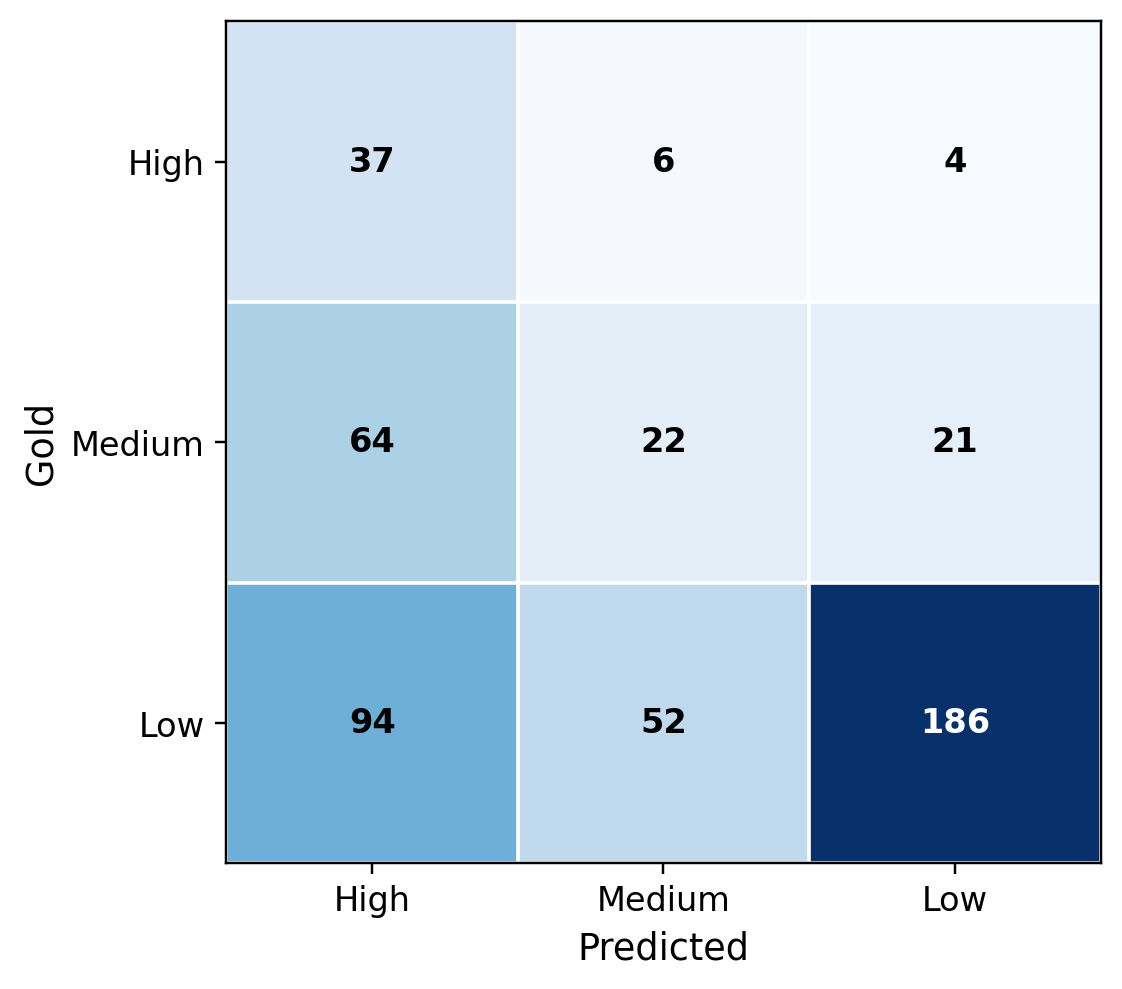}
}
\subfloat[Gemma-12B (Zero-shot)]{
  \includegraphics[width=0.24\textwidth]{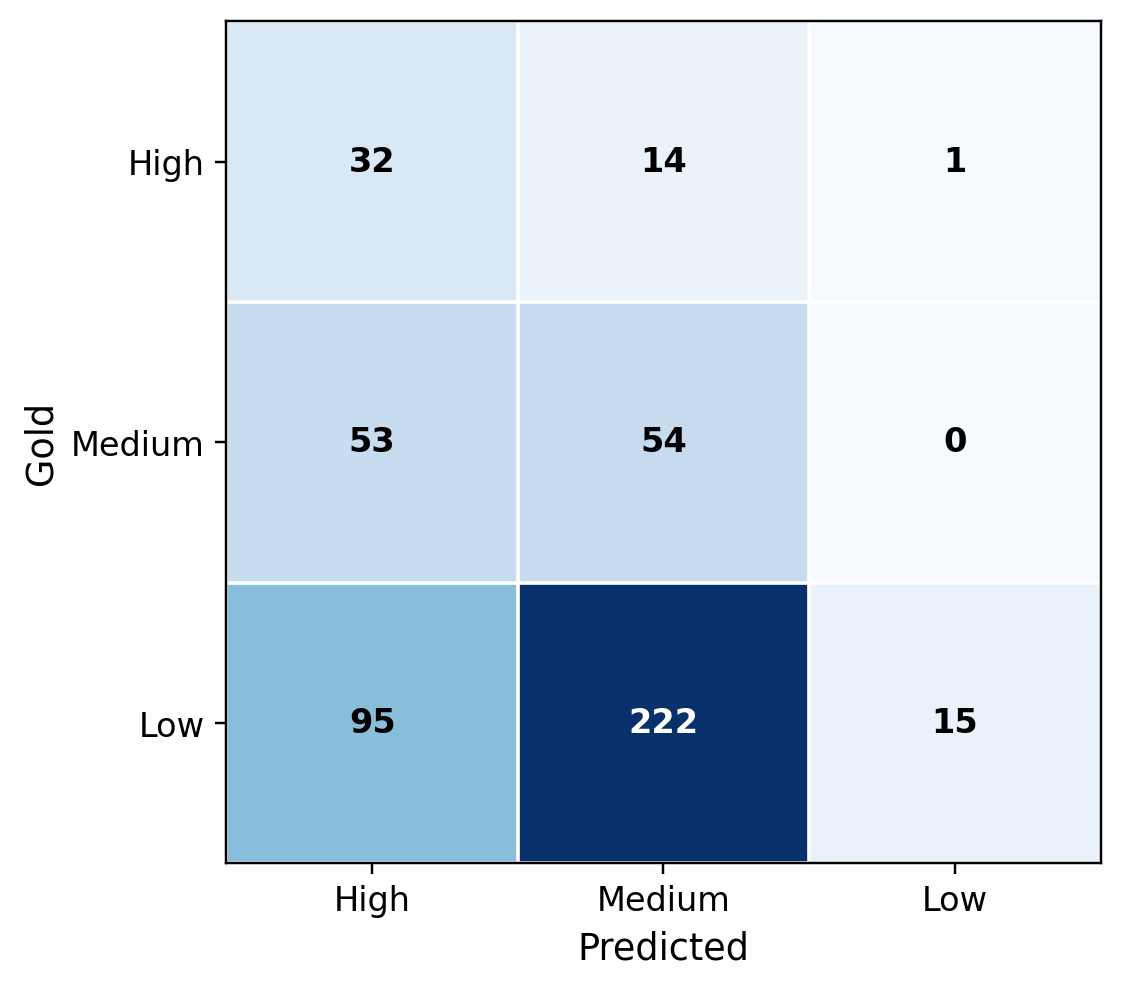}
}
\subfloat[Gemma-12B (Fine-tuned)]{
  \includegraphics[width=0.24\textwidth]{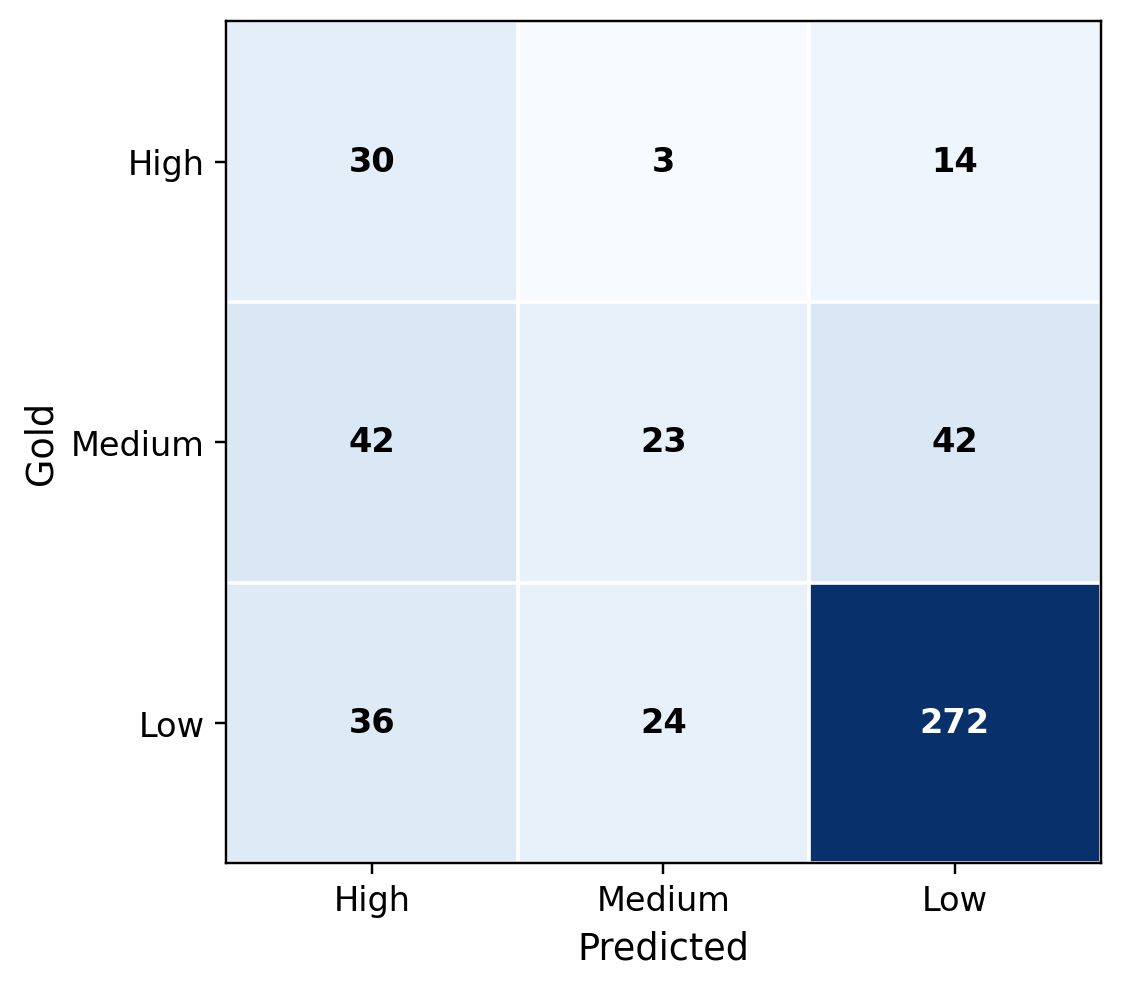}
}

\caption{Confusion matrices for selected models before (zero-shot) and after fine-tuning.}
\label{fig:confusion_matrices}
\end{figure*}

\section{Discussion}
\label{sec:discussion}

\subsection{Overview of the results}
Instruction-tuned LLMs, when evaluated in a zero-shot setting, demonstrate limited effectiveness in assessing the severity of threat events. Most models do not surpass a simple majority-class baseline, highlighting that threat-level determination is a non-trivial task requiring task-specific supervision. A primary contributing factor is the systematic bias of the models toward overestimating severity, often predicting \textit{High} or \textit{Medium} levels even for low-risk events. This behavior may be due to the alignment with general safety-oriented heuristics that encourage conservative predictions without task-specific guidance. 

Figure~\ref{fig:confusion_matrices} presents confusion matrices for selected models before and after fine-tuning. In the zero-shot setting, predictions are generally biased toward higher threat levels. In contrast, fine-tuned models generate a distribution of predictions that aligns more closely with the actual class proportions in the test set, where \textit{Low} represents the majority class. 

As expected, larger models generally perform better, but the effect is not strictly linear. In the zero-shot setting (Table~\ref{tab:leaderboard_base}), Command-R+ (111B) achieves the highest macro-F1 (0.313), followed by Llama-3.1-8B (0.277) and Llama-3.3-70B (0.267).

After fine-tuning on the balanced dataset (Table~\ref{tab:leaderboard_finetuned_64}), the larger models achieve the highest performance: Llama-3.3-70B reaches a macro-F1 score of 0.574, followed by Command-R+ at 0.554 and Llama-3.1-8B at 0.520. Interestingly, the gap between the 70B and 8B models shrinks to just over 5 points, showing that task-specific supervision significantly reduces the performance advantage of larger models. Smaller models benefit most in absolute terms, with Gemma-3-1B gaining +0.350 F1 and Gemma-3-12B +0.288. 

These findings indicate that while increasing model size can offer incremental gains, fine-tuning on task-specific data is the dominant factor for performance, suggesting that smaller, well-tuned models can approach the effectiveness of much larger models with considerably lower computational cost. Ultimately, the difficulty lies in the inherently subjective and operational nature of threat-level determination, which extends beyond pattern recognition and into domain expertise. To capture the broader context, LLMs must be able to trace the full progression of a cybersecurity incident, from an isolated malicious network activity to its potential escalation into a large-scale attack. Incorporating notions of causality into the training process may help LLMs better model these dynamics and improve their ability to assess threat level.

\subsection{Effect of Input Representation}

Input format has a consistent impact across all settings (Table~\ref{tab:delta_md_json}). Switching from JSON to the textual representation \textit{Text} format improves macro-F1 in every case. In the zero-shot setting, gains range from +0.027 (Llama-3.3-70B) to +0.184 (Gemma-3-12B). \textit{Text} also improves accuracy, likely due to fewer formatting errors and better alignment with the prompt. After fine-tuning, the same trend holds: all models see macro-F1 gains with textual input. The two Gemma models are again notable, where textual representation improves macro-F1 while reducing accuracy, suggesting less over-prediction of the majority class. This means that models are able to produce more balanced predictions across all threat levels. These findings highlight that how MISP events are represented is a central design choice in LLM-based CTI classification, where concise formatting of inputs can directly enhance robustness.

\begin{table}[t]
\centering
\caption{The comparison of input representations. Positive values indicate that the textual representation achieves a better score than JSON.}
\label{tab:delta_md_json}
\begin{tabular}{lcc}
\hline
Model & Base & Fine-tuned  \\
\hline
\hline
Model & $\Delta (Acc./F1)$ & $\Delta  (Acc./F1)$   \\
\hline
Llama-3.3-70B        & +0.028 / +0.027 & +0.035 / +0.028 \\
Command-R+ (111B)    & +0.109 / +0.106 & +0.009 / +0.037 \\
Llama-3.1-8B         & +0.066 / +0.049 & +0.035 / +0.040 \\
Gemma-3-12B-IT       & +0.163 / +0.184 & -0.177 / +0.196 \\
Gemma-3-1B-IT        & +0.090 / +0.056 & -0.237 / +0.125 \\
\hline
\end{tabular}
\end{table}

\subsection{Fine-Tuning Procedure} \label{sec:how-to-finetune}

While fine-tuning yields substantial performance improvements, its effectiveness depends strongly on the composition of training data. Our initial experiments revealed that training on the full dataset, without accounting for class imbalance, resulted in overfitting, even for large models. Models fine-tuned on the full set exhibited a strong bias toward the dominant \textit{Low} class, predicting it in over 90\% of test cases regardless of input. This resulted in deceptively high accuracy but poor macro-F1, as minority classes were consistently under-predicted.

To address this, we opted to fine-tune on a down-sampled variant of the dataset with equal instances per class (111 per threat level). This balanced setup produced more robust and generalizable models, leading to substantial macro-F1 improvements across all architectures.

We also explored the effect of LoRA configurations by conducting a limited hyper-parameter search on two representative models: Llama-3.1-8B and LLama-3.3-70B. Specifically, we evaluated three common settings: $(r=8, \alpha=32)$, $(r=16, \alpha=64)$, and $(r=32, \alpha=128)$. Based on performance on a held-out 20\% development split, we found that $(r=16, \alpha=64)$ consistently offered the best performance. Due to resource constraints, this hyperparameter search sweep was only conducted for the two models mentioned above. The remaining models were fine-tuned directly using the $(16, 64)$ setting on the balanced training set.

 \subsection{Comparison with other CTI Tasks}

By comparing our results with those reported in CTIBench \cite{alam2024ctibench}, we see that the tasks evaluated in CTIBench generally achieve slightly higher performance than threat-level determination, even though performance still varies across tasks and models. In our case, Llama-3.3-70B and Llama-3.3-8B without fine-tuning reach only 0.23–0.30 in accuracy and macro-F1. In contrast, the same models obtain clearly stronger results in CTIBench: according to Table 1 of their study, accuracy across three tasks (CTI-MCQ, CTI-RCM, CTI-TAA) exceeds 0.36, with Llama-3.3-70B reaching a macro-F1 of 0.47. This difference suggests that threat-level determination, while highly relevant for operations, is a more difficult task and requires further adaptation and research before it can deliver deployable results. Finally, CTIBench results show ChatGPT-4 as the strongest model overall, whereas in our threat-level evaluation, no single model dominates; after fine-tuning, however, Llama-3.3-70B stands out as the best performer.

\section{Conclusion and Future Work}
\label{sec:conclusion}
In conclusion, this paper provides a narrow and deep evaluation of LLMs for the task of threat level determination in cyber threat intelligence. Our benchmark, built on curated MISP OSINT feeds, reveals that while zero-shot LLMs exhibit limited effectiveness, supervised fine-tuning significantly enhances their performance, achieving macro-F1 scores between 0.40 and 0.58. These results highlight that task-specific adaptation is essential for leveraging LLMs in CTI applications. Nevertheless, the overall performance remains insufficient for reliable operational deployment, underscoring the need for further research focused on improving data quality, optimizing model architectures, and developing domain-specific tuning strategies to fully harness LLMs in threat intelligence workflows.

As future work, we plan to expand the dataset by extracting CTI information from a broader range of threat intelligence reports, beyond OSINT feeds, to capture more prosperous and diverse adversarial behaviors. This information can be integrated into a structured knowledge graph, which would not only facilitate more effective representation of complex CTI relationships but also serve as a foundation for fine-tuning LLMs. By combining knowledge graph–based reasoning with LLM adaptability, we aim to enhance the accuracy and interpretability of automated threat-level determination.

\section{Acknowledgement}
This work was supported in part by the Horizon Europe project HARPOCRATES (GA No. 101069535) and CUSTODES (GA No. 101120684).

We acknowledge the use of AI generative tools to assist with language refinement and improving the clarity of the sections of this paper, while ensuring that the core scientific contributions remain original and the result of our independent research.

\bibliographystyle{IEEEtran}
\bibliography{references.bib}

\end{document}